\documentclass[prl,twocolumn,showpacs,amsmath,amssymb,superscriptaddress]{revtex4}
\usepackage{graphicx}% Include figure files
\usepackage{graphics}% Include figure files
\usepackage{dcolumn}% Align table columns on decimal point
\usepackage{bm}% bold math
\begin{document}

\title{Quantum-tomography of entangled photon pairs by quantum-dot cascade 
decay}

\author{F. Troiani}
\affiliation{Departamento de F\'{\i}sica Te\'orica de la Materia
Condensada, Universidad  Aut\'onoma de Madrid, Cantoblanco 28049
Madrid, Spain}
\affiliation{CNR-INFM National Research Center on nanoStructures and 
bioSystems at Surfaces (S3), 41100 Modena, Italy}

\author{J.I. Perea} 
\affiliation{Departamento de F\'{\i}sica Te\'orica de la Materia
Condensada, Universidad  Aut\'onoma de Madrid, Cantoblanco 28049
Madrid, Spain}

\author{C. Tejedor}
\affiliation{Departamento de F\'{\i}sica Te\'orica de la Materia
Condensada, Universidad  Aut\'onoma de Madrid, Cantoblanco 28049
Madrid, Spain}

\begin{abstract}
We compute the concurrence of the polarization-entangled photon pairs 
generated by the biexciton cascade decay of a semiconductor quantum dot. 
We show how a cavity-induced increase of the photon rate emission
reduces the detrimental effect of the dot dephasing and of the excitonic 
fine structure. However, strong dot-cavity couplings and finite detection 
efficiencies are shown to reduce the relevance of the desired cascade 
decay with respect to that of competing processes. 
This affects the merits of the entangled photon-pair source, beyond 
what estimated by the quantum-tomography. 
\end{abstract}

\pacs{78.67.Hc, 42.50.Dv, 03.67.Hk}

\maketitle

Most protocols in optical quantum-information processing require deterministic 
sources of entangled photon pairs~\cite{nielsen}. 
It has been argued that semiconductor quantum dots (QDs) might represent the 
active element of such a quatum device~\cite{benson,stace}. 
In fact, first proofs-of-principle have been recently established, where  
entanglement between the polarization and frequency degrees of freedom has 
been measured in photon pairs generated by the cascade emission from single 
QDs~\cite{akopian,stevenson06}.  
However, the degree of entanglement is still limited by the dot dephasing, 
the excitonic fine-structure splitting, and the mixed nature of the dot state, 
resulting from its incoherent (i.e., off-resonant) excitation.
It is believed that these limitations can be to a large extent overcome by 
increasing the photon rate emission, through the embedding of the QD in a 
semiconductor microcavity (MC)~\cite{young06}. The developing of more 
sophisticated devices and excitation startegies is also being 
considered~\cite{benson,fattal04_3,benyoucef04}. 
In spite of such great interest, a clear theoretical interpretation 
of the recent experimental achievements is not currently available. 
It is the goal of the present paper to provide such an understanding,
and the resulting indications for the future development of 
entangled-photon sources. 

\begin{figure}[tbp]
\includegraphics[width=\columnwidth]{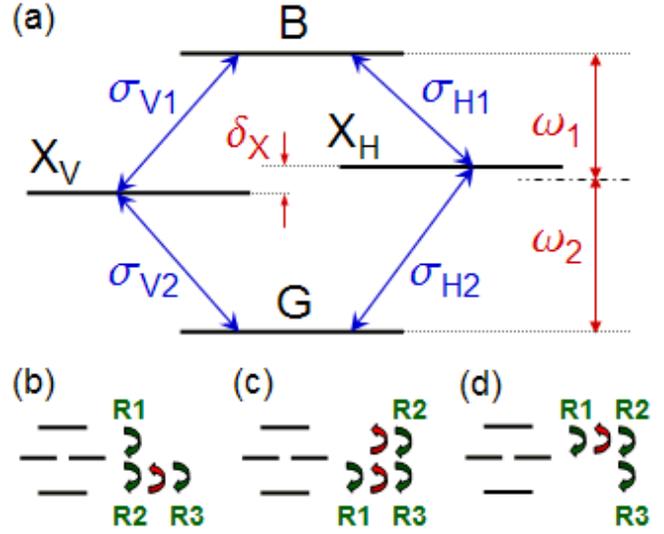}
\caption{(a) Level scheme of the QD, including the ground state $G$, the 
linearly polarized excitons $X_V$ and $X_H$, and the biexciton $B$. 
The optical transitions between them are induced by photons with linear 
polarization, frequencies $\omega_n \pm \delta_X / 2$ ($n=1,2$),   
and are represented by the ladder operators
$\sigma_{H1} = |X_H\rangle\langle B|$, 
$\sigma_{H2} = |G\rangle\langle X_H|$, 
$\sigma_{V1} = |X_V\rangle\langle B|$ and
$\sigma_{V2} = |G\rangle\langle X_V|$. 
(b-d) Examples of excitation (red) and relaxation (green) sequences, leading 
to the emission of at least one photon from the decay of the biexciton and one 
from that of an exciton.} 
\label{fig1}
\end{figure}

The origin of the polarization-frequency entanglement in the photon pair
resides in the QD's low-energy level scheme (Fig.~\ref{fig1}(a)).
This includes the biexciton state ($B$), the two lowest exciton levels ($X_H$ 
and $X_V$), and the ground state ($G$). 
After the system is excited to state $ | B \rangle $, it radiatively decays 
through a sequential emission of two photons, with either horizontal ($H$) 
or vertical ($V$) linear polarizations. 
The photons generated by the biexciton and exciton decays (1 and 2, respectively), 
differ in frequency due to the biexciton binding energy 
$\Delta =\omega_2 -\omega _1$. 
In the ideal case, the dot is initially driven to $ | B\rangle $, and 
subsequently relaxes, generating the maximally entangled two-photon state $
| \psi \rangle = ( |H1,H2\rangle + |V1,V2\rangle ) / \sqrt{2} $
by cascade emission. 
In realistic exciting conditions, however, the state of the emitted radiation 
$\rho_{ph}$ is affected by a number of uncertainties; these include 
the number of emitted photons and the photon-emission time 
(time jitter)~\cite{kiraz04,troiani06}.
The quantum-tomography experiments are based on coincidence measurements, where
one projects $\rho_{ph}$ onto the two-photon subspace spanned by the basis 
$\{ |H1,H2\rangle , |H1,V2\rangle , |V1,H2\rangle , |V1,V2\rangle \} 
$~\cite{akopian,stevenson06}. 
The corresponding two-photon density matrix is
\begin{eqnarray}
\rho_{ph}^{QT} = \left(
\begin{array}{cccc}
\alpha   &     0 &     0 & \gamma \\
     0   & \beta &     0 &      0 \\
     0   & 0     & \beta &      0 \\
\gamma^* &     0 &     0 & \alpha
\end{array}
\right) , \label{rho_ph}
\end{eqnarray}
where all the coherences but that between $|H1,H2\rangle$ and
$|V1,V2\rangle$ are identically zero. 
While the ideal state
$|\psi\rangle$ corresponds to setting $\beta = 0$ and $\gamma =
\alpha = 1/2$, the presence of imperfections and the same exciting
conditions cause departures from ideality, resulting in $ | \gamma |
< \alpha $ and $ \beta > 0$. 
Here, we shall be concerned with both the degree of 
entanglement of $\rho^{QT}_{ph}$ and with its overlap with the overall 
radiation state $\rho_{ph}$. This determines the quality of a 
dot-based source of entangled photon pairs, beyond what estimated by 
the quantum tomography.

\begin{figure}[tbp]
\begin{center}
\includegraphics[width=\columnwidth]{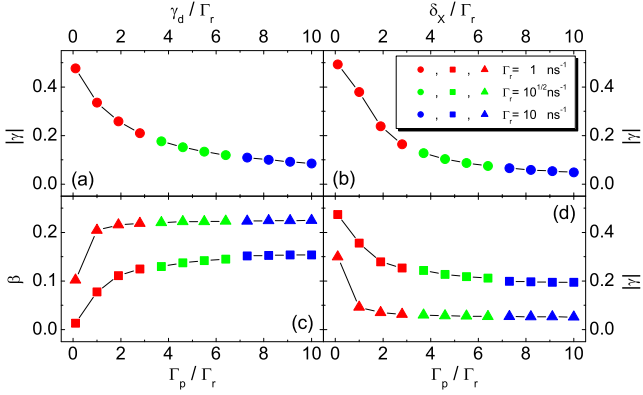}
\caption{Values of $ \beta $ and $ |\gamma |$ (see Eq.~\ref{rho_ph}) as a 
function of the dephasing rate $\gamma_d$ (a), of the energy splitting 
$\delta_X$ (b), and of the pumping rate $\Gamma_p$ (c,d).
In the upper panels the QD is initialized to the biexciton state, with 
$\delta_X = 0$ (a) or $\gamma_d = 0$ (b);
in the lower panels, it is continuously pumped, with 
$ |\psi (0)\rangle = |G\rangle $ and $\delta_X = \gamma_d = 0 $.
In panels (a,b) $t_m = t_m' = 0$ and $t_M = t_M' = 5/\Gamma_r $; in (c,d)
$t_m = t_M = t_\infty$, $t_m' = 0$, and $t_M' = 1/\Gamma_r$ (triangles) 
or $ t_M' = 5/\Gamma_r$ (squares).}
\label{fig2}
\end{center}
\end{figure}

The state of the emitted radiation can be derived from that of the dot-cavity 
system ($\rho$). The time evolution of $\rho$ is computed by solving the 
following master equation, within the Born-Markovian and
rotating-wave-approximations~\cite{perea05,troiani06} ($\hbar =1$):
\begin{equation}
\dot\rho = i [\rho , H] 
+ (\Gamma_r\mathcal{L}_{QD}^{r} 
+  \Gamma_p\mathcal{L}_{QD}^{p} 
+  \gamma_d\mathcal{L}_{QD}^{d} 
+  \kappa\mathcal{L}_{MC}) \, \rho .
%\label{mastereq}
\nonumber
\end{equation}
Here 
$ H_{int} = H-H_{QD} - H_{MC} = \sum_{\zeta=H,V} \sum_{n}  
( g_n \sigma_{\zeta n} a_\zeta^\dagger + \mathrm{H.c.} ) $ 
is the dot-cavity interaction Hamiltonian, 
with 
$ \sigma_{\zeta 1} \equiv | G       \rangle\langle X_\zeta | $, 
$ \sigma_{\zeta 2} \equiv | X_\zeta \rangle\langle B       | $,
and $a_\zeta$ the cavity destruction operators.
The QD Liouvillian includes the contribution from the radiative relaxation
($\Gamma_r\mathcal{L}_{QD}^{r}$), that from the incoherent pumping 
($\Gamma_p\mathcal{L}_{QD}^{p}$), and the term accounting for the pure 
dephasing ($\gamma_d\mathcal{L}_{QD}^{d}$). 
Finally, the coupling of the MC with the external modes and the resulting
photon-loss process are accounted for by  
$ \kappa \mathcal{L}_{MC} $.

The emission process produces radiation in a mixed state, from which
the two-photon coincidence measurements single-out 
the $\rho_{ph}^{QT}$. Its matrix elements, experimentally
reconstructed~\cite{benson,fattal04_3,benyoucef04,akopian,stevenson06,young06} 
by means of the tomographic method~\cite{james,altepeter}, 
theoretically correspond to a specific set of (time-averaged) second-order 
correlation functions:
%\begin{widetext}
\begin{eqnarray}
\alpha &=&
%\langle HH |\rho_{ph}| HH \rangle = 
A\int_{t_m}^{t_M} \int_{t_m'}^{t_M'} \; dt \, dt' \; \langle
\,\sigma^{\dagger}_{H1} (t) \,\sigma^{\dagger}_{H2} (t')
\,\sigma_{H2} (t')
\,\sigma_{H1} (t) \rangle
\nonumber \\
\beta &=&
%\langle HV |\rho_{ph}| HV \rangle = 
A\int_{t_m}^{t_M}
\int_{t_m'}^{t_M'} \; dt \, dt \; \langle
\,\sigma^{\dagger}_{H1} (t) \,\sigma^{\dagger}_{V2} (t')
\,\sigma_{V2} (t')
\,\sigma_{H1} (t) \rangle
\nonumber \\
\gamma &=& 
%\langle HH |\rho_{ph}| VV \rangle =
A \int_{t_m}^{t_M}
\int_{t_m'}^{t_M'} \; dt \, dt' \; \langle
\,\sigma^{\dagger}_{H1} (t) \,\sigma^{\dagger}_{H2} (t')
\,\sigma_{V2} (t') \,\sigma_{V1} (t) \rangle
\label{correlationfunct}
\end{eqnarray}
%\end{widetext}
where the normalization constant $A$ is such that $2(\alpha +
\beta)=1$. 
The correlation functions appearing in Eq.~\ref{correlationfunct} 
are computed by applying the quantum regression theorem; this results in a 
set of equations analogous to those that apply to the evolution of $\rho$.

A number of criteria have been proposed to establish wether or not a given 
$\rho$ is separable. 
According to the Peres separability criterium~\cite{peres}, $\rho_{ph}^{QT}$ 
is entangled if and only if $ | \gamma| > \beta $. 
In the case of a two-qubit system, the concurrence~\cite{james} ($C$) also
allows to quantify the degree of entanglement. 
In this specific case, it is easily shown that 
$ C = 2 (|\gamma|-\beta) $ for $ | \gamma| > \beta $, 
and $ C = 0 $ otherwise. 
Three main effects degrade the ideal (maximally entangled)
$\rho_{ph}^{QT}$ to a separable one: {\it (i)} the pure dephasing
affecting the QD tends to quench the phase coherence of the
intermediate state resulting from the first photon emission, $ | X_H
\rangle \otimes | H1 \rangle + | X_V \rangle \otimes | V1 \rangle $,
and therefore to reduce $\gamma$; {\it (ii)} the energy splitting $
\delta _X $ between the two excitonic states suppresses the
interference terms by providing a which-path information (reduced $
\gamma $); {\it (iii)} the contribution to $\rho_{ph}^{QT}$ of photons
generated in different cascade emissions (see, e.g., Fig.~\ref{fig1}(b-d)) 
results both in a finite probability of observing counter-polarized 
1 and 2 photons ($\beta > 0 $) and in that of loosing phase coherence 
between the H and V components of each photon type ($| \gamma |< \alpha $). 
For the sake of clarity, we start by considering these effects separately.

In order to isolate the contribution of the pure dephasing {\it (i)}, 
we set $\delta _X = 0 $ and initialize the QD to the biexciton state, 
$ |\psi (0) \rangle = |B\rangle $, in the absence of pumping terms
($\Gamma_{p}=0$). In Fig.~\ref{fig2}(a) we plot $ | \gamma | $ as a
function of the dephasing rate $ \gamma_d $, normalized to the
emission rate $ \Gamma_r $. Due to the absence of an
excitation source, there is no probability for the QD of being
re-excited after emission, and therefore for the $ \rho_{ph}^{QT} $ to
suffer from the mixing of different cascades. As a consequence,
$\beta = 0$ and the Peres criterion is trivially satisfied by any
$\gamma \neq 0$. 
The fact that the points descibe a single curve (i.e., that $C$ depends
on $ \gamma_d $ and $ \Gamma_r $ only through their ratio) provides a 
clear evidence of the interplay between dephasing and photon rate emission: 
in fact, a fast emission of photon 2 reduces the time during 
which dephasing degrades the intermediate state of the dot-cavity 
system~\cite{benson}.
{\it (ii)} The effect of the energy splitting $\delta _X$ is shown in 
Fig.~\ref{fig2}(b), where we plot $| \gamma | $ as a function of 
$ | \delta _X | / \Gamma_r $, with $ \gamma_d =0$. 
Once again, $| \gamma |$ and $ C = 2 |\gamma |$ depend on the two parameters 
only through their ratio. 
In fact, an increased $ \Gamma_r $ results in an enhanced intrinsic 
line-width of the excitonic transitions, and therefore increases the 
overlap between the wave-packets corresponding to photons $H2$ and $V2$. 
{\it (iii)} The possibility that $ \rho_{ph}^{QT} $ may
include contributions from different cascades arises from the 
finite probability of re-exciting the system between the first and the
second photon emission. 
This explains the increment of $ \beta $ and fall of $ \gamma $ for 
increasing excitation rate, shown in Fig.~\ref{fig2}(c,d) 
for the case of a continuously pumped QD (with $ | \psi (0)
\rangle = |G\rangle $ and $ \delta_X = \gamma_d = 0$). 
In this case, an importan role is also played by the time interval associated 
with the detection of photon 2 ($ \Delta t' \equiv t_M' - t_m' $). 
A shorter $ \Delta t' $ reduces the probability, e.g., that the detected photons
1 and 2 might arise respectively from the relaxations $R1$ and $R3$ in 
Fig.~\ref{fig1}(b). However, while this allows to increase the concurrence of 
$ \rho_{ph}^{QT} $, it doesn't improve the merits of the entangled-photon 
source, which depend on the state of the overall emission $ \rho_{ph} $ (see below). 

The above results allow to isolate the contribution of different physical 
effects to the degradation of the ideal $ \rho_{ph}^{QT} $, and thus provide 
upper limits to the value of $C$ corresponding to each $\gamma_d$, $\Gamma_p$,
or $\delta_X$. 
The incidence of each of these factors strongly depends on the emission rate
of photon 2. 
Therefore, in the following we shall analyze the case where such emission rate 
is increased by embedding the QD in a semiconductor MC close to resonance with 
the excitonic transition ($ \omega_c = \omega_2 $). 
Besides, we shall focus on the case of pulse-pumped 
excitation~\cite{stevenson06,akopian,young06}, for it allows to trigger the 
generation of photon pairs and reduces the probability of unwanted re-excitations 
of the dot after the first cascade. 
In the weak-coupling regime, the effect of the QD-MC interaction essentially 
consists in enhancing the photon-emission rate by a factor corresponding to 
the so-called Purcell factor, $F_p = 2g^2 / \kappa \Gamma_r $. 
The contribution to the parameters $ \alpha $, $ \beta $, and $ \gamma $ arising 
from the cavity emission are computed by replacing in Eq.~\ref{correlationfunct}
the ladder operators $ \sigma_{\zeta 2} $ with $ a_\zeta $ ($\zeta = H,V$).

\begin{figure}[tbp]
\begin{center}
\includegraphics[width=\columnwidth]{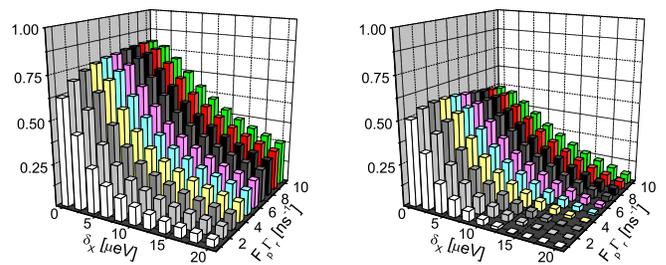}
\caption{Concurrence of $ \rho_{ph}^{QT}$, as a function of the energy
splitting $\delta _X$ and the effective emission rate of the second
photon $ F_p \Gamma_r $. Results are shown for the case of a short gaussian
pulse ($\sigma = 10\, $ps, $\Gamma_p = 0.05\, $ps$^{-1}$, left panel) and for 
a long one ($\sigma = 100\, $ps, , $\Gamma_p = 0.005\, $ps$^{-1}$, right panel).
Other physical parameters are: $\Gamma_r^{-1} = 10^3\,$ps, $\gamma_d^{-1} = 
500\, $ps, $g_{H,V} = 0.05\,$meV.} 
\label{fig3}
\end{center}
\end{figure}

In Fig.~\ref{fig3} we plot the values of the concurrence as a function of the 
energy splitting $\delta_X$ and of the effective emission rate $F_p\Gamma_r$ of 
the excitonic transition. 
For a relatively short exciting pulse (gaussian time profile, with $\sigma = 
10\, $ps) and $ \delta_X $ not larger than a few $ \mu eV $ (left panel), 
values of $C$ of about 0.8 are achieved. 
While the dependence of $C$ on $\delta_X$ is monotonous, that on the Purcell
factor is characterized by the presence of a maximum. 
In fact, if $ F_p \Gamma_r $ is too high with respect to the exciting rate 
$\Gamma_p$, the system tends to emit a photon from an excitonic level before 
being excited to state $B$ (see below). 
The fact that the photons 1 and 2 might procede from 
different cascades (e.g., from the decays $R2$ and $R1$, respectively, in 
Fig.~\ref{fig1}(c)), weakens the overall degree of polarization correlation.
This feature is even more dramatic in the case of a large and weaker exciting 
pulse (right panel, $\sigma = 100\, $ps), where it results in a strong 
suppression of the concurrence and in a displacement of the maximum towards
lower values of $F_p$.
Therefore, while increasing the Purcell factor allows to suppress the 
detrimental effects of dephasing and of the exciton energy splitting $\delta_X$ 
(see Fig.~\ref{fig2}), large values of $F_p$ result in an overall reduction of 
the frequency-polarization entanglement.

A good source of single entangled photon pairs is one where the probability $p$
of emitting only two photons from a single cascade 
($ B \rightarrow X \rightarrow G $)
is high as compared to those of the competing processes
(Fig.~\ref{fig1}(b-d)).  
In order to estimate such probabilities and to gain a deeper understanding on
the underlying processes, it is useful to distinguish between the properties of 
the radiation emitted by the dot-cavity system and those of the detected 
photons.
This calls for including the quantum feedback of the continuous measurement on 
the dot-cavity system~\cite{wiseman}.
In particular, the evolution conditioned upon not having detected any photon 
up to time $t$ is obtained by applying to the Liouvillians $\mathcal{L}_{QD}^r$
and $\mathcal{L}_{MC}$ the following substitutions:
\begin{eqnarray}
\mathcal{L}_{MC}\, \rho
& \longrightarrow & 
\mathcal{L}_{MC}\, \rho - \eta_{\scriptscriptstyle{MC}}
\sum_{\zeta = H,V} \, a_\zeta \, \rho \, a^\dagger_\zeta ,
\nonumber \\
\mathcal{L}_{QD}\, \rho
& \longrightarrow & 
\mathcal{L}_{QD}\, \rho - 
\eta_{\scriptscriptstyle{QD}}
\sum_{\zeta = H,V} \sum_{n=1,2}
\, \sigma_{\zeta n} \, \rho \, \sigma_{\zeta n}^\dagger ,
\label{conditional}
\end{eqnarray}
where $\eta_{\scriptscriptstyle{QD}}$ ($\eta_{\scriptscriptstyle{MC}}$)
is the efficiency of the detectors times the collection efficiency of the 
photons emitted by the cavity (dot). 
The probability that the first detected photon is generated with polarization 
$\zeta$, by the relaxation 
of the dot into the leaky modes or by the cavity loss, is given respectively by 
\begin{eqnarray}
p_{\zeta n}^{(\eta )} & = & \Gamma_r \eta_{\scriptscriptstyle{QD}} 
\int dt \, \langle 
\sigma^\dagger_{\zeta n} (t) \sigma_{\zeta n} (t) 
\rangle_{\eta} \, ,
\nonumber\\
p_{\zeta c}^{(\eta )} & = & \kappa \eta_{\scriptscriptstyle{MC}} 
\int dt \, \langle 
a^\dagger_{\zeta} (t) a_{\zeta} (t) 
\rangle_{\eta} \, ,
\end{eqnarray}
where $ \zeta = H , V $, $ n = 1 , 2 $, and 
$\eta = (\eta_{\scriptscriptstyle{QD}},\eta_{\scriptscriptstyle{MC}})$.
If also the second detection is taken into account, the corresponding joint 
probabilities take the form
\begin{eqnarray}
p^{(\eta )}_{\zeta 1, \xi c} & \!\!\! = \!\!\! &
B
\int_{t_m}^{t_M} \!\!\!  dt  \int_{t}^{t_M} \!\!\! dt' \, 
\langle \,
\sigma_{\zeta 1}^{\dagger} (t) \, 
a_{\xi}^{\dagger} (t') \, a_{\xi} (t') 
\, \sigma_{\zeta 1} (t) 
\, \rangle_{\eta} ,
\nonumber\\
p^{(\eta )}_{\zeta 1, \xi 2} & \!\!\! = \!\!\! &
B'
\int_{t_m}^{t_M} \!\!\!  dt  \int_{t}^{t_M} \!\!\! dt' \, 
\langle \,
\sigma_{\zeta  1}^{\dagger} (t) \, 
\sigma_{\xi    2}^{\dagger} (t') \, \sigma_{\xi  2} (t') 
\, \sigma_{\zeta 1} (t) 
\, \rangle_{\eta} ,
\nonumber
\end{eqnarray}
where $ B  = \kappa   \eta_{\scriptscriptstyle{MC}} \, 
             \Gamma_r \eta_{\scriptscriptstyle{QD}} $,
      $ B' = (\Gamma_r \eta_{\scriptscriptstyle{QD}})^2 $,
and $\xi = H,V$.
If all the emitted photons are detected 
($ \eta_{\scriptscriptstyle{QD}} = \eta_{\scriptscriptstyle{MC}} = 1 $), 
the above quantities reflect the intrinsic properties of the photon source. 

\begin{figure}[tbp]
\begin{center}
\includegraphics[width=\columnwidth]{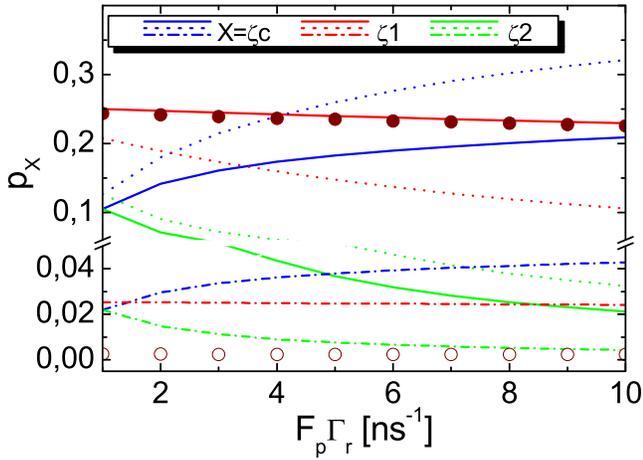}
\caption{Probabilities $p_X^{(\eta )}$ associated with the first photon detection,  
as a function of the Purcell factor $F_p$ ($\delta_X=0$). 
The probabilities are computed for short ($\sigma = 10\, $ps, $\Gamma_p = 0.05\, 
$ps$^{-1}$, solid and dot-dashed lines) and long ($\sigma = 100\, $ps, $\Gamma_p = 0.005\, 
$ps$^{-1}$ dotted lines) laser-pulse durations, for high ($\eta_{\scriptscriptstyle{QD}} = 
\eta_{\scriptscriptstyle{MC}} = 1$, solid and dotted) and low 
($\eta_{\scriptscriptstyle{QD}} = \eta_{\scriptscriptstyle{MC}} = 0.1$, 
dotted-dashed) detection efficiencies.
The filled (empty) circles correspond to 
$ p_{\zeta 1, \zeta 2}^{(\eta )} + p_{\zeta 1, \zeta c}^{(\eta )} $ in the 
high (low) efficiency case, being $\zeta = H, V$.} 
\label{fig4}
\end{center}
\end{figure}

In Fig.~\ref{fig4} we show the dependence of $ p_{\zeta c}^{(\eta )} $ and 
$ p_{\zeta n}^{(\eta )} $ on the Purcell factor (that on $ \delta_X $, not shown here,
is negligible).
For $\sigma = 10\,$ps (solid lines) and with growing $F_p$, 
$ p_{\zeta c}^{(1,1)} + p_{\zeta 1}^{(1,1)} $
slightly increases with respect to $p_{\zeta 2}^{(1,1)}$. 
Correspondingly, the probability of generating only an entangled photon pair,
$ p^{(\eta )} \le p_{H 1}^{(\eta )} + p_{V 1}^{(\eta )} $, 
decreases below 0.46 for $F_p = 10$.
This effect is more evident in the long-pulse case (dotted 
lines), where the exciton relaxation rate $ (F_p +1)\Gamma_r $ becomes larger than the 
excitation rate $\Gamma_p$.
% The number of cases where no photons at all are emitted, quantified by
% $ p_0^{(1,1)} = {\rm Tr} 
% [ \rho ]_{(1,1)} 
% \simeq 0.0082 $, is approximately independent of $F_p$, and 
% does not change with $\sigma$ (given that the exciting
% intensity $ \Gamma_p $ times the pulse duration coincide in the two cases).
The analysis of the second photon detection shows that, in the short-pulse case,  
$ p_{\zeta 1}^{(1,1)} \simeq p_{\zeta 1, \zeta 2}^{(1,1)} + 
p_{\zeta 1, \zeta c}^{(1,1)} $, while the consecutive emission of two 
photons from the $B$ state is highly improbable,
$ p_{\zeta 1, \xi 1}^{(1,1)} \leq 0.0025 $.
Therefore the probability of generating only the required photon pair 
is well approximted by that of emitting from the biexciton state first, 
$ p^{(1,1)} \simeq p^{(1,1)}_{H 1} + p^{(1,1)}_{V 1} $.
Things change qualitatively when imperfections in the photon detections are
taken into account 
($ \eta_{\scriptscriptstyle{QD}} = \eta_{\scriptscriptstyle{MC}} = 0.1 $,
dotted-dashed lines).
In fact, besides the order-of-magnitude reduction of all the detection 
probabilities, the observed weight of the single cascade $ B \rightarrow X 
\rightarrow G $ is reduced with respect to that of the undesired processes:
$ p^{(0.1,0.1)}_{\zeta 1} < p^{(0.1,0.1)}_{\zeta c} + p^{(0.1,0.1)}_{\zeta 2} $
and 
$ p^{(0.1,0.1)} \ll p^{(0.1,0.1)}_{H 1} + p^{(0.1,0.1)}_{V 1} $.
Therefore, an improved detection efficiency increases not only the fraction 
of useful excitation cycles, but also the degree of entanglement in the 
observed photon pairs. 

In conlusion, the coupling of the QD with a MC and the resulting increase 
of the photon-emission rate compensate the effect of dephasing and of the 
exciton energy splitting on the entanglement of the emitted photon pairs.
However, large Purcell factors also reduce the probability of the desired 
cascade decay with respect to that of competing processes, resulting in an 
overall decrese of the concurrence. 
Finally, due to the finite detection efficiency, the observed degree of 
polarization correlation, is smaller than that of the emitted photons.
Such limitations, as well as those related to the time-jitter, might be 
possibly overcome by coherently exciting the QD with two-photon absorption 
processes~\cite{akimov06}.

This work has been partly supported by the Spanish MEC under
contracts MAT2005-01388, NAN2004-09109-C04-4, by CAM under
Contract S-0505/ESP-0200, and by the Italy-Spain "integrated 
action" HI2005-0027.

%%%%%%%%%%%%%%%%%%%%%%%%%%%%%%%%%%%%%%%%%%%%%%

\bibliography{paper}

%%%%%%%%%%%%%%%%%%%%%%%%%%%%%%%%%%%%%%%%%%%%%%

\end{document}